\documentclass[pra,twocolumn,floatfix,a4paper,superscriptaddress,showkeys,superscriptaddress,longbibliography]{revtex4-1}
\usepackage{bm,color,graphicx,amsmath,txfonts,tikz}
\usepackage{siunitx}%sistema SI
\usepackage[caption=false]{subfig}
\usepackage{hyperref}%collegamenti ipertestuali
\usepackage[english]{babel}%per scrivere in inglese	
\usepackage{subfig}
\usepackage{datetime}

\newcommand\hlighty[1]{\tikz[overlay, remember picture,baseline=-\the\dimexpr\fontdimen22\textfont2\relax]\node[rectangle,fill=yellow,rounded corners,fill opacity = 0.2,draw,thick,text opacity =1] {$#1$};}%Highlight single matrix elements, yellow

\newcommand\hlightv[1]{\tikz[overlay, remember picture,baseline=-\the\dimexpr\fontdimen22\textfont2\relax]\node[rectangle,fill=blue!50,rounded corners,fill opacity = 0.2,draw,thick,text opacity =1] {$#1$};}%Highlight single matrix elements, violet
\usetikzlibrary{fit}%highlight block matrices
\newcommand{\tikzmark}[2]{\tikz[overlay,remember picture,
  baseline=(#1.base)] \node (#1) {#2};}
\newcommand{\Hlightr}[1][submatrix]{%
    \tikz[overlay,remember picture]{
    \node[rectangle,fill=red,rounded corners,fill opacity = 0.2,draw,thick,text opacity =1,inner sep=0pt,fit=(left.north west) (right.south east)] (#1) {};}
}%red
\newcommand{\Hlightg}[1][submatrix]{%
    \tikz[overlay,remember picture]{
    \node[rectangle,fill=green,rounded corners,fill opacity = 0.2,draw,thick,text opacity =1,inner sep=0pt,fit=(left.north west) (right.south east)] (#1) {};}
}%green
\usetikzlibrary{arrows,matrix,positioning}
\newcommand{\ee}{{\rm e}}

\newtheorem{defn}{Definition}
\begin{document}

\title{Master thesis: High-rate multipartite quantum secret sharing with continuous variables}

\author{Jacopo Angeletti}

\begin{abstract}
Quantum cryptography has undergone substantial growth and development within the multi-disciplinary field of quantum information in recent years. The field is constantly advancing with new protocols being developed, security measures being improved, and the first practical applications of these technologies being deployed in optical fibers and free space optical beams. In this paper, we present a comprehensive review of a cutting-edge metropolitan-scale protocol for continuous-variable quantum cryptography. The protocol allows an arbitrary number of users to send modulated coherent states to a relay, where a generalised Bell detection creates secure multipartite correlations. These correlations are then distilled into a shared secret key, providing a secure method for quantum secret-sharing. This novel approach to quantum cryptography has the potential to offer high-rate secure multipartite communication using readily available optical components, making it a promising advancement in the field.
\end{abstract}


\maketitle
\formatdate{7}{4}{2020}

\section{Introduction}
Quantum key distribution (QKD)~\cite{Pirandola_2020, RevModPhys.74.145} with continuous-variable (CV) systems~\cite{e17096072} has garnered significant attention in recent years. The design of CV-based QKD protocols utilizing Gaussian quantum states of optical beams has proven to be particularly effective, and these states can now be easily produced in laboratory settings. The ideal implementation of QKD protocols that utilize CV systems~\cite{RevModPhys.77.513, Serafini2017} and Gaussian states~\cite{RevModPhys.84.621} has the potential to approach the PLOB bound~\cite{Pirandola2017,PhysRevLett.102.050503}, which is the ultimate limit of point-to-point communication. These advancements demonstrate the exciting progress and potential for continued development in the field of QKD with CV systems. Recently, there has been a significant push towards an end-to-end approach that can be applied to network implementations~\cite{PhysRevA.99.030301, PhysRevA.91.022320, Ottaviani2019}. This approach utilizes an intermediate relay as a means of communication, allowing parties to perform measurement-device-independent (MDI) QKD protocols~\cite{PhysRevA.91.022320, Pirandola2015}, even if the relay is untrusted. This development provides a solution that can greatly benefit network implementations and has garnered significant attention in the field.

We analyze a cutting-edge multipartite protocol for secure quantum secret-sharing (QSS) that utilizes CV systems and an MDI configuration. This protocol can be easily implemented using linear optics and provides a secure method for key distribution. In this protocol, an arbitrary number of users are divided into groups and send Gaussian-modulated coherent states to an untrusted relay. A generalized multipartite Bell detection is performed at the relay and the results are publicly broadcast. QSS enables the distribution of a secret key among all users, which requires their collaboration for validity. In the case of non-collaboration, a threshold behavior is manifested and allows for the detection of ``dummy" users, leading to the potential abort of the protocol. This multipartite protocol based on CV systems and MDI configuration provides a promising solution for secure key distribution in a network setting.

Consider a configuration where users are distributed asymmetrically around a relay station and analyze the security of the protocol against collective attacks. In this scenario, we assume that Eve uses independent entangling cloners~\cite{Pirandola2017, Grosshans_2003} and analyze the asymptotic regime of many (ideally infinite) exchanged signals. The links connecting the parties to the relay are modeled as memory-less thermal-loss channels, with the assumption that users in the same ensemble share both common transmissivity and thermal noise. Under these realistic conditions, we demonstrate that the protocol is suitable for metropolitan-scale areas. For example, the ultimate limit for bipartite secure communication still allows for the establishment of a secret key between two groups in a noisy environment within a radius of $10$ km.

The paper is organized as follows: in Sec.~\ref{descrip}, we describe the communication scheme. Sec.~\ref{res2} focuses on the analysis of bi-partitions of users for a thermal-loss channel. In Sec.~\ref{res3}, we examine two specific configurations, referred to as the $Y$- and $X$-schemes, which allow for secure secret-sharing among three and four groups, respectively. Finally, in Sec.~\ref{secConcl}, we summarize our findings and provide concluding remarks. To facilitate a deeper understanding of the protocol, the mathematical tools used in our analysis are provided in the appendices.

\section{Description of the communication scheme}\label{descrip}
\begin{figure*}
    \centering
    \subfloat[\centering]{{
        \includegraphics[width=6cm]{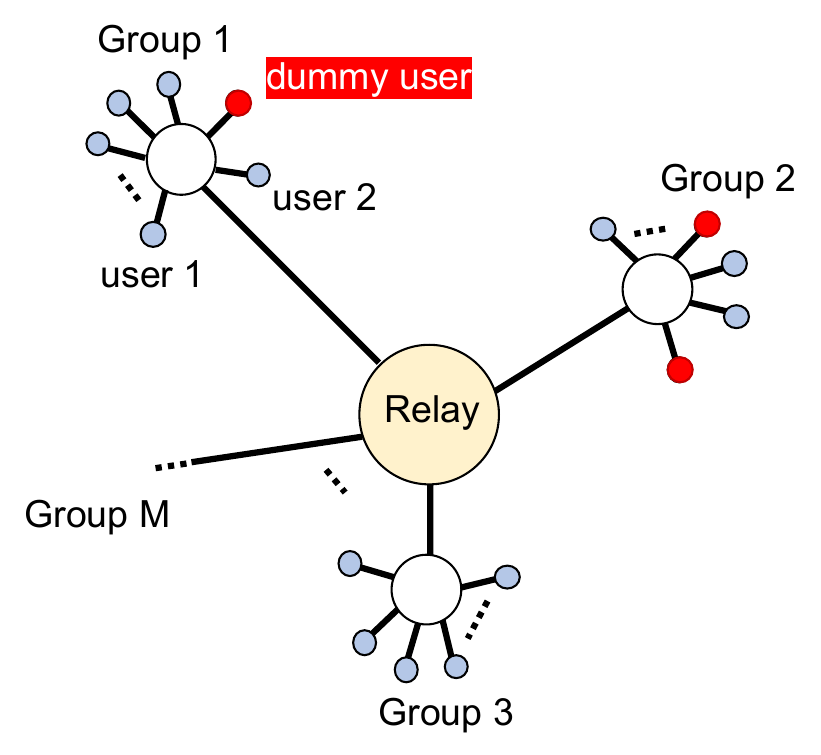} 
        \label{dummy_scheme}
    }}%
    \qquad
    \subfloat[\centering]{{
        \includegraphics[width=10cm]{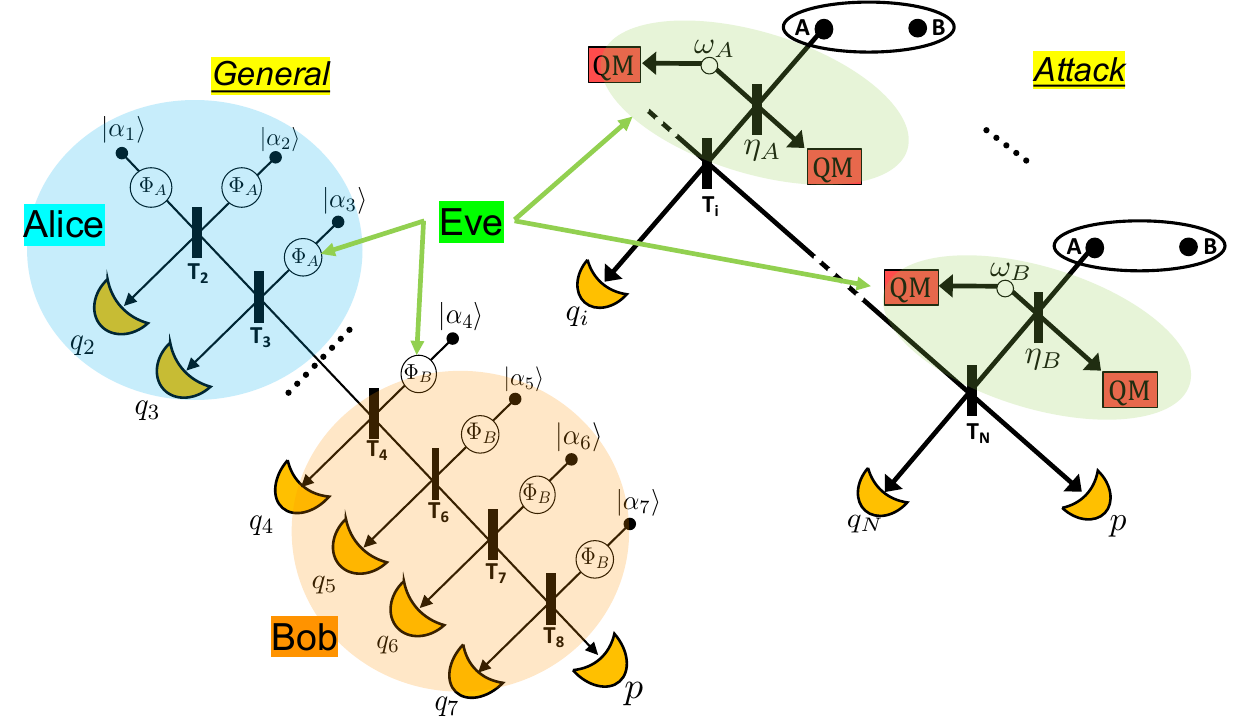} 
        \label{qssscheme5}
    }}%
    \caption{
   (a) Illustration of a group of $N$ users organized into $M$ ensembles, each consisting of $N_{j}$ users, where $j=\left\{1,\,\dots,\,M\right\}$, such that $\sum_{j=1}^MN_j\leq N$. The possible presence of ``dummy" users is represented in red. The ensembles are arranged at different distances from the relay, while the users within each ensemble are at equal distances from the relay (Note: the illustration may not be entirely accurate in terms of distances between ensembles and relay). (b) \textbf{(General)} Prepare and measure (PM) implementation of the QSS scheme for $N=7$ users, with $N_1=3$ in group ``$1$", and $N_2=4$ users in group ``$2$". It may be generalised to an arbitrary number $M$ of groups, each with a different number $N_j$ of users users. Each user, referred to as ``Bob," sends a Gaussian-modulated coherent state with amplitude $\left|\alpha_k\right\rangle$ to an untrusted relay through an optical link described by a thermal-loss channel $\Phi_j$. At the relay, the incoming states undergo a generalised multipartite Bell detection, performed through a cascade of beamsplitters and homodyne detectors. The beamsplitters have transmissivities $T_1=1$ and $T_k=1-k^{-1}$ for $k=\left\{2,\,\dots,\,N\right\}$, while the homodyne detectors measure either the $\hat q$ or the $\hat p$ quadrature, as described in the figure. The outcome, $\gamma:=\left(p,\,q_2,\dots,\,q_N\right)$ is broadcast to the Bobs, so that \textit{a posteriori} correlations are created among their local variables $\alpha_1,\,\dots,\,\alpha_N$. These correlations are used to extract a secret key for QSS. \textbf{(Attack)} Sample of the protocol in the EB representation, where each thermal-loss channel $\Phi_j$ is characterised by its transmissivity $\eta_j$ and thermal noise $\omega_j=2\bar n_j+1$.
    }
\end{figure*}
We provide the definition of a generic secret-sharing protocol as follows:
\begin{defn}
	An $(M,N)$-threshold scheme is a procedure for dividing a message into $N$ pieces, called shadows or shares, such that no subset of fewer than $M$ shadows can reveal the message, but any set of $M$ shadows can be used to reconstruct it~\cite{PhysRevA.59.1829}.
 \end{defn}
To illustrate this concept, consider the scenario of Alice setting up a launch program for a nuclear warhead from a remote location. To ensure that the launch cannot be initiated by a single person, she divides the launch code into $N$ parts, and distributes them among $N$ individuals. These shares are encrypted and contain no information about the original launch code individually. However, if $M$ individuals cooperate, they would be able to reconstruct the complete launch code. This makes it more challenging for any single person to gain unauthorized access, as they would need to collude with $M-1$ others.

In order to perform a QSS protocol, consider an arbitrary number $N$ of trusted users (referred to as ``Bobs") arranged into $M$ groups, with $N_{j}$ users in each group, where $j=\left\{1,\,\dots,\,M\right\}$. The sum of all users in the groups should not exceed $N$ (see Fig.~\ref{dummy_scheme}), and when $\sum_{j=1}^MN_j=N$, we refer to this as the ``full-house" case. The users send random Gaussian-modulated coherent states $\left|\alpha_k\right\rangle$ through a thermal-loss channel $\Phi_j$ to an untrusted relay, where a generalized multipartite Bell detection is performed, as depicted in Fig.~\ref{qssscheme5}. The relay is modeled as an $N$-port interferometer consisting of $N$ beamsplitters, with increasing transmittivities $T_1=1$ to $T_k=1-k^{-1}$ for $k=\left\{2,\,\dots,\,N\right\}$, followed by $N$ homodyne detections. The first output is measured in $\hat p$, while the rest are $\hat q$-homodyned, where $\hat q$ and $\hat p$ are the two quadrature operators of the optical mode such that $\left[\hat q,\,\hat p\right]=2i$. The outcome $\gamma:=\left(p,\,q_2,\dots,\,q_N\right)$ is broadcast to all Bobs, who can then remove the local displacement caused by the measurements. Further mathematical details are provided in App.~\ref{appInterf}.

The theoretical assessment of the protocol is performed in the entanglement-based (EB) representation. In this representation, each source of coherent states is represented by a two-mode squeezed vacuum (TMSV) state $\hat\rho_{AB}$, which undergoes heterodyne detection. The $\hat B$ modes are kept at each user's station, while the $\hat A$ modes are sent to the relay for detection. As a result, each user is equipped with a TMSV state $\hat\rho_{AB}$ that has a zero mean and a covariance matrix (CM) that is equal to
\begin{equation}\label{VABTMSV}
	\textbf{V}_{AB}=
		\left(
		\begin{array}{cc}
			\mu\textbf{I}	 &\sqrt{\mu^2-1}\textbf{Z} 	\\
			\sqrt{\mu^2-1}\textbf{Z} &\mu\textbf{I}  	\\ 	
		\end{array}
		\right),
\end{equation}%
where $\textbf{Z}=\text{diag}\left\{1,\,-1\right\}$, $\textbf{I}=\text{diag}\left\{1,\,1\right\}$, $1\leq \mu:=\cosh 2r \in \mathbb R$~\footnote{We sometimes omit that $\mu$ and $\omega$ are in SNU for simplicity of notation. This is also due to the fact that it is usually not possible to carry out a dimensional check in this field.}, and the modes are ordered as $\left(\hat q^{A}, \hat p^{A}, \hat q^{B}, \hat p^{B}\right)^T$. Here, $r$ is the squeezing parameter. By heterodyning mode $\hat B$, each Bob remotely prepares a coherent state $\left|\beta\right\rangle$ on mode $\hat A$, the amplitude of which is modulated by a complex Gaussian with variance $\mu - 1$. For large modulation $\mu\gg 1$, the outcome of the measurement $\widetilde\beta\simeq\alpha^*$ is approximately equal to the projected amplitude $\alpha$.
\begin{figure}[ht]
	\begin{center}
  		\includegraphics[width=\columnwidth, scale=2]{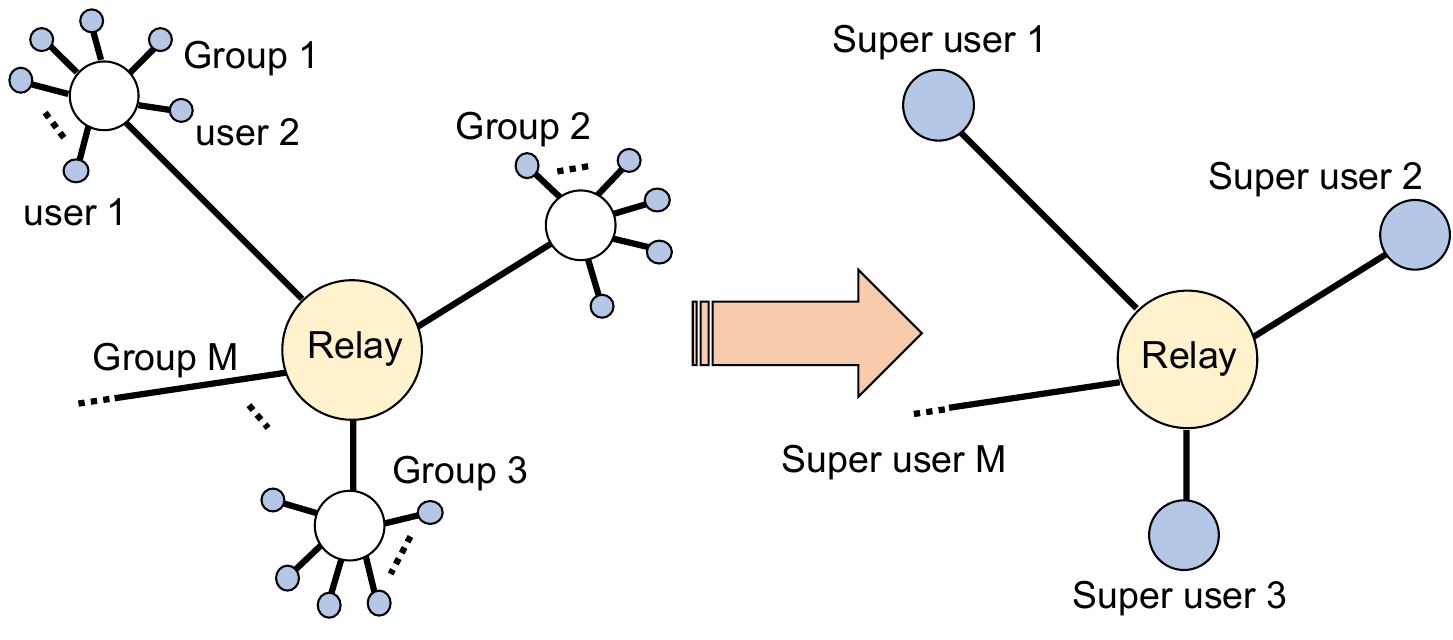}
		\caption{Unitary entanglement localisation in $M$-symmetric states. Within each group, users cooperate to concentrate the entanglement they share, and we can describe the situation from the point of view of $M$ ``super users," which are the $M$ groups of users.}
		\label{reductionfig}
	\end{center}
\end{figure}
The CM of the TMSV state $\hat\rho_{AB}$ Eq.~(\ref{VABTMSV}), upon the action of the channel $\Phi_j$, undergoes the transformation 
\begin{equation}\label{VpABch}
	\textbf{V}'_{AB}=
		\left(
		\begin{array}{cc}
			x_j\boldsymbol I	 &z_j \boldsymbol Z \\
			z_j \boldsymbol Z	 &y\boldsymbol I \\
		\end{array}
		\right),
\end{equation}%
with $j=\left\{1,\,\dots,\,M\right\}$. Here, each thermal-loss channel $\Phi_j$ is characterized by its transmissivity $\eta_j$ and thermal noise $\omega_j$, such that
\begin{equation}\label{xyzdef}
    \begin{split}
        x_j&=\eta_j\mu+\left(1-\eta_j\right)\omega_j,\\
        y&=\mu,\\
        z_j&=\sqrt{\eta_j\left(\mu^2-1\right)}.
    \end{split}
\end{equation}
After the Bell measurement and communication of the outcome $\gamma$, the modes $\hat{\textbf B}:= \hat{B}_1\cdots \hat{B}_N$ are projected onto a symmetric $N$-mode Gaussian state (see also App.~\ref{bell}). The users are divided into $M$ groups, each consisting of $N_j$ members, and the global state is represented by $\rho_{\textbf M|\gamma}$, where $\hat{\textbf M}:= \hat{N}_1\cdots \hat{N}_M$ represents all the members of the $M$ groups. The members of each group can apply local operations (LOs)~\footnote{In the EB representation, these LOs can be implemented by means of suitable interferometers, one on each side. These passive LOs can be available also at the post-processing stage, after the action of the relay, for the equivalent PM description.} on $\rho_{\textbf M|\gamma}$ to establish a common secret key among the $M$ groups. These local Gaussian operations concentrate the quantum correlations of all the Bobs, transforming $\rho_{\textbf M|\gamma}$ into an effective $M$-mode Gaussian state $\rho_{M|\gamma}$ with CM~\footnote{Across the $M$-partition, plus a tensor product of thermal states for the remaining modes.}~\cite{PhysRevA.71.032349}
\begin{equation}\label{VMgamma}
	\textbf{V}_{M|\gamma}=
		\left(
		\begin{array}{cccc}
			\boldsymbol \Gamma_{11}	 &\boldsymbol \Gamma_{12}&\cdots&\boldsymbol \Gamma_{1M} \\
			\boldsymbol \Gamma_{21}	 &\boldsymbol \Gamma_{22}&\cdots&\boldsymbol \Gamma_{2M} \\
			\vdots &\vdots &\ddots &\vdots\\
			\boldsymbol \Gamma_{M1}	 &\boldsymbol \Gamma_{M2}&\cdots&\boldsymbol \Gamma_{MM}
		\end{array}
		\right),
\end{equation}%
where~\footnote{The index $k$ always runs from $1$ to $M$, but we explicitly show it only once for cleanliness.}
\begin{equation}\label{gammaij}
	\begin{split}
			\boldsymbol \Gamma_{ij}&=y \textbf I\,\delta_{ij}\\
		&-z_iz_j\textrm{diag}\left(\frac{\delta_{ij}\sum_{k\neq i}N_k\mathfrak X^{(ki)}-\left(1-\delta_{ij}\right)\mathfrak X^{(ij)}}{\sum_{k\neq i}N_k\mathfrak X^{(i)}},\,\frac{\sqrt{N_iN_j}}{\sum_{k=1}^M N_kx_k}\right),
	\end{split}		
\end{equation}%
with $\delta_{ij}$ the Kronecker delta, and
\begin{equation}
	\mathfrak X^{(\alpha\beta)}:=\prod_{k\neq\alpha\neq\beta} x_k.
\end{equation}%
As a result, we can consider the situation from the perspective of $M$ aggregated entities, commonly referred to as ``super users," which correspond to the $M$ groups into which the users are divided. This is shown in Fig.~\ref{reductionfig}. For more information on this process, refer to App.~\ref{reduction}.

We consider the practical limitations that arise during the implementation of the multipartite Bell detection. The presence of inefficiencies in the detectors is accounted for by including detector efficiencies $\tau<1$ in the homodyne measurements. This is achieved through the use of $N$ beam splitters with transmissivity $\tau$. In CV-Bell detections, it is possible to attain high efficiencies at both optical and telecom frequencies, both with and without fiber components. Despite the technical difficulties, homodyne detection can reach detection efficiencies of up to $90\%$~\cite{PhysRevLett.68.3020, PhysRevA.67.033802, Grosshans_2003}.

To account for the finite effects due to a limited number of exchanged signals between the parties, we must consider the reconciliation efficiency $\xi$ of the classical codes used for error correction and privacy amplification~\cite{Pirandola_2020, RevModPhys.74.145}. Despite being crucial for extracting a secret key, this process typically has an efficiency $\xi<1$, with typical values ranging from $\xi\simeq 0.95\div 0.985$~\cite{PhysRevA.77.042325, doi:10.1142/S0219749915500100, Milicevic2018, 10.5555/3179575.3179579, Wang2018, Gehring2015}. In addition to this, there may be other imperfections that arise from the relay, such as the asymmetric behavior of the interferometer beamsplitters~\cite{Ottaviani2019}. However, this case is nontrivial and requires numerical solutions, and is therefore not considered in this analysis.

Assuming asymptotic security and infinite Gaussian modulation~\footnote{In order to reach the optimal and asymptotic performances provided by the infinite-dimensional Hilbert space.}, the secret-key rate of the protocol against collective attacks is simply given by~\cite{Pirandola_2020}
\begin{equation}\label{erreibgammachiegamma}
	R=\xi I_{\textbf B|\gamma}-\chi_{E|\gamma}.
\end{equation}%
For practical purposes, the secret-key rate must be optimized over the modulation parameter $\mu$, as outlined in App.~\ref{appRate}. To analyze the potential performance of the protocol, we will focus on scenarios that are suitable for experimental testing, by considering the cases with $M=\left\{2,\,3,\,4\right\}$.

\section{Results}

\subsection{Bipartite system}\label{res2}
\begin{figure}[ht]
	\begin{center}
  		\includegraphics[width=.9\columnwidth]{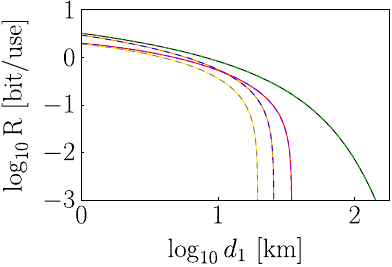}
		\caption{Secret-key rates (in bits per use) for two optimally distributed groups, as a function of the distance $d_1$ (in km, for a standard optical fiber) of group ``$1$", with group ``$2$" fixed at the relay ($d_2=0\,\si{\kilo\meter}$). The different curves correspond to different parameter settings: black ($\omega_1=\omega_2=\tau=1$), dashed green ($\omega_1=1,\,\omega_2=1.1,\,\tau=1$), red ($\omega_1=\omega_2=1,\,\tau=0.98$), dashed purple ($\omega_1=1,\,\omega_2=1.1,\,\tau=0.98$), orange ($\omega_1=1.1,\,\omega_2=1=\tau=1$), dashed blue ($\omega_1=\omega_2=1.1,\,\tau=1$), yellow ($\omega_1=1.1,\,\omega_2=1,\,\tau=0.98$), and dashed gray ($\omega_1=\omega_2=1.1,\,\tau=0.98$). The performances of the protocol, whether ideal (black and orange curves) or not (red and yellow curves), are not significantly affected by the presence of noise in the nearest group (dashed green, blue, purple, and gray curves). However, in the case of thermal noise corresponding to $\omega_1=1.1$ SNU and a detection efficiency of $\tau=98\%$, the performance is reduced by about $150\%$.}
		\label{Best2}
	\end{center}
\end{figure}%
In a QSS session, users are divided into $M=2$ groups, referred to as group ``$1$" and group ``$2$", with group ``$2$" positioned deeper~\footnote{Consisting the interferometer of a cascade of beam splitters $T_k$, we may define a user depth. A user, whose channel is characterised by $T_i$, is deeper than another $\left(T_{j}\right)$ if $i>j$.} in the interferometer and serving as the decoder~\footnote{We are actually inverting the roles of encoder and decoder between the groups with respect to Refs.~\cite{Pirandola2017, Pirandola2015}. This will end up with a trivial and irrelevant exchange of roles between Alice and Bob.} (see also App.~\ref{bipartsys}). As shown in Fig.~\ref{Best2}, the performance of the protocol is evaluated in terms of the secret-key rate, measured in bits per channel use, as a function of the distance $d_1$ between group ``$1$" and the relay (measured in kilometers for a standard optical fiber with $0.2$ dB$/$km attenuation). The relay is fixed at a distance of $d_2 = 0\, \si{\kilo\meter}$ from group ``$2$". The notation used is the following 
\begin{defn}
	A splitting of the kind ``X/Y" means that $X\%$ ($Y\%$) of all users belongs to group ``$1$" (``$2$").
\end{defn}%
In Fig.~\ref{Best2}, we compare the optimal rate for different channel types (thermal- and pure-loss) and detection efficiencies, with group ``$2$" fixed at the relay and group ``$1$" at varying distances. The parameters of the thermal-loss channel are detailed in the figure caption. Our Gaussian QSS scheme achieves outstanding performance compared to qubit-based protocols, with secret key rates that are at least three orders of magnitude higher~\cite{PhysRevLett.108.130503}, over comparable distances, for which one has $\lesssim 10^{-4}$ bit$/$use at $\lesssim 25$ km. If symmetric configuration was limited to $3.8$ km, our apparatus can reach a maximum distance of $170$ km in standard optical fibers, with a high key rate of $2\times10^{-4}$ bit/use. With a clock of $25$ MHz, this corresponds to a key rate of the order of $2.5$ Mbits$/$sec for all users. The optimal bipartition corresponds to the ``full-house" case and symmetric splitting, resulting in the same best (ideal) performance as standard CV-MDI-QKD~\cite{Pirandola2017, Pirandola2015} in its asymmetric configuration (black curve). This is possible because the state $\rho_{2|\gamma}$ is independent from the number of users $N$. Our results show that, while imperfections and noise do have a general destructive effect, they do not appreciably affect the performance of the protocol (solid and dashed line coincide). However, the presence of thermal noise with $\omega_1=1.1$ SNU and a detection efficiency of $\tau=98\%$ reduces the performance by about $150\%$. These results highlight the feasibility of high-rate secure CV-MDI-QKD QSS in a noisy environment at a metropolitan scale.

\subsubsection{Secret-key rate versus group distance}
\begin{figure}[ht]
	\includegraphics[width=.9\columnwidth]{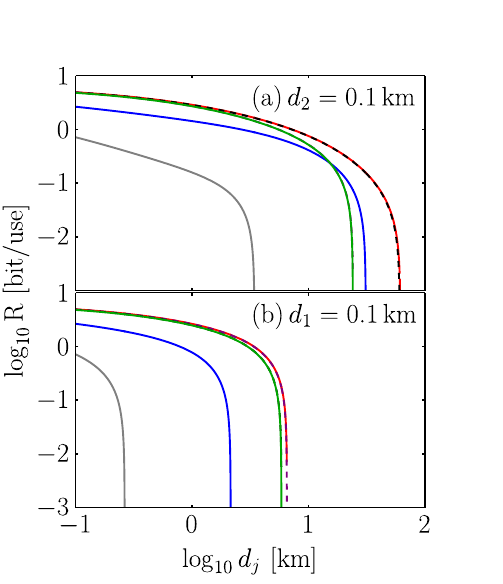}
	\caption{performance of the protocol in terms of secret-key rate (in bits per use) as a function of the distance of one of the groups (in km, for a standard optical fiber), with the distance of the other group fixed at $d_{i}=0.1\,\si{\kilo\meter}$. Figures $a$ and $b$ differ in which group is fixed. The different curves correspond to different splitting ratios: $50/50$ (red, black, green, and purple), $5/95$ (blue), and $1/99$ (gray). We also consider different amounts of noise added to the optimal $50/50$ configuration, with noise parameters of $\omega_1=\omega_2=1$ (red, blue, and gray), $\omega_1=1,\,\omega_2=1.1$ (black), $\omega_1=1.1,\,\omega_2=1$ (purple), and $\omega_1=\omega_2=1.1$ (green). The plots are invariant with respect to asymmetrical splittings when there is no injection of noise, implying no depth effects induced by the relay. The best performance is achieved for the optimal $50/50$ splitting, with the deepest group in the interferometer closest to the relay, resulting in an improvement of a factor of $10$ compared to the symmetric protocol~\cite{Ottaviani2019}. The ultimate limit for bipartite secure communication still allows for the establishment of a secret key between two groups at a metropolitan scale within a radius of $d_2=\si{10\,\kilo\meter}$ (not shown). Noise is more tolerable in the nearest group, where the performance is not significantly affected.}
	\label{rdpanel1}
\end{figure}%
In this study, we examine the behavior of the secret-key rate with respect to the distance of one of the two groups, while the other is fixed. We vary the distance $d_i=\left\{0.1\,\si{\kilo\meter},\,1\,\si{\kilo\meter},\,10\,\si{\kilo\meter}\right\}$ of one of the groups from the relay. We focus on the full-house case for different splittings, including $50/50$, $5/95$, and $1/99$. The $50/50$ splitting is also analyzed in a noisy environment.

The trends~\footnote{We decide just to expose the results concerning $0.1\,\si{\kilo\meter}$ being the others totally analogous.} in the secret-key rate behavior with respect to the distance of a group are presented in Fig.~\ref{rdpanel1}. Our analysis shows that for a pure-loss channel, the performance is worse when the splittings are extreme and does not vary~\footnote{A protocol is invariant with respect to asymmetrical splittings if they lead to the same result.} with asymmetrical splittings. This indicates that there are no depth effects induced by the relay. Despite this, we are pleased to find that the ultimate limit for bipartite secure communication still allows for the establishment of a secret key between two groups within a radius of $10$ km in a metropolitan scale (not shown). When studying the impact of noise, we observe that in the case of asymmetrical noise~\footnote{That is, when only one of the two groups is affected by noise (we exclude the case in which both are, clearly the worst possible scenario).}, the group closer to the relay can tolerate more noise in its link, and the performance is not significantly affected. At present, reasonable values of excess noise are in the range of $\epsilon=0.04\div 0.05$ SNU~\cite{Zhang_2019}, which express the incredible tolerance of our protocol to noise, despite the conversion from excess noise to thermal noise $\omega$ not being immediate.

\begin{figure}[ht]
	\centering
		\includegraphics[width=.9\columnwidth]{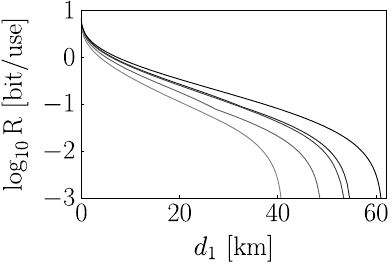}
		\caption{
  Secret-key rates (bits per use) are plotted against the distance $d_1$ (in km, for a standard optical fiber) of the superficial group ``$1$" in a two-group setup, where group ``$2$" is fixed at a distance of $d_2=0.1\,\si{\kilo\meter}$. The groups are optimally distributed in a $50/50$ ratio. The figure shows the impact of reconciliation efficiency on the secret-key rates, with different curves representing different values of reconciliation efficiencies. Moving from right to left, the values are $\xi=1$, $0.985$, $0.98$, $0.95$, and $0.90$. As expected, the secret-key rate decreases with decreasing reconciliation efficiency. The results highlight the importance of high reconciliation efficiency to achieve a higher secret-key rate in quantum key distribution protocols.
  }
		\label{rateRE}
\end{figure}

Finally, Fig.~\ref{rateRE} further (see Fig.~\ref{Fig5}) illustrates the impact of distance on the secret-key rate when considering non-ideal reconciliation efficiency ($\xi \leq 1$). As expected, the rate decrease as the distance increases, and the impact of imperfect reconciliation becomes more pronounced. The results clearly demonstrate the need for efficient reconciliation to achieve high secret-key rates over large distances.

\subsubsection{Threshold behaviour}
\begin{figure}[ht]
	\includegraphics[width=.9\columnwidth]{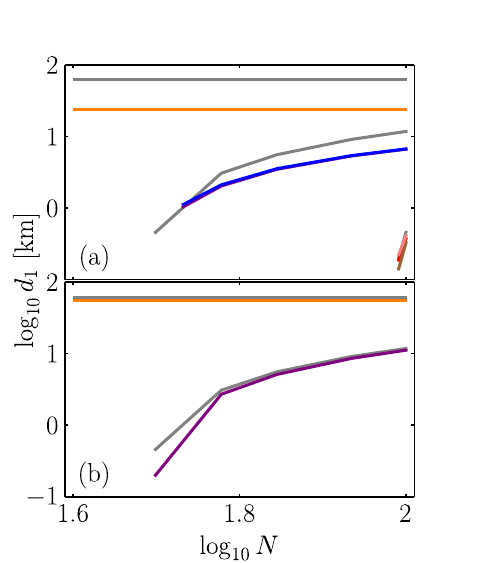}
	\caption{Maximum fiber distance for the QSS protocol, plotted for three different types of bipartitions of the users. The optimal $50/50$ splitting (orange curves) exhibits performance that is independent of $N$. We consider the scenario with one dummy Bob per group, i.e., $N_1 = N/2 - 1$ and $N_2 = N/2$ (purple curves), or vice versa, $N_1 = N/2$ and $N_2 = N/2-1$ (blue), and two dummy Bobs, $N_1 = N_2 = N/2 - 1$ (red), or $N_1=N/2-2$ and $N_2=N/2$ (brown), or vice versa, $N_1=N/2$ and $N_2=N/2-2$ (pink). The protocol performance is always the worst when the users in the shallowest group do not cooperate. In (a), we compare the performance with the corresponding pure-loss channel (gray). In (b), we consider the case of reconciliation efficiency $\xi= 0.985$ for the worst-case scenario of one dummy Bob in group ``$1$". Any configuration of two dummy users produces a negative rate. The introduction of reconciliation efficiency lowers the curves and makes the threshold behavior more pronounced when compared with the ideal $\xi=1$ case (gray).}
	\label{threshbplots}
\end{figure}%
The results of this study, depicted in Fig.~\ref{threshbplots}, showcase the threshold behavior characteristic of QSS. As depicted, when one or more users do not cooperate, the performance drops significantly, which allows for easy detection and potential termination of the session.

Our study focuses on determining the maximum distance achievable by one group of users, $d_j^{\,max}$~\footnote{We omit the superscript max in the figures for simplicity of sketching.}, as a function of the number of users $N$, while keeping the distance of another group fixed at $d_i$. Three different types of user bipartitions were considered. As previously stated, the optimal split of $50/50$ (represented by orange in the figure) has a performance that does not depend on $N$. Additionally, we analyze the cases where one ``dummy" user is present in each group, leading to two scenarios: $N_1 = N/2 - 1,\,N_2 = N/2$ (purple) and $N_1 = N/2,\,N_2 = N/2-1$ (blue).

We also considered the scenario where two dummy users are present, resulting in three possible combinations: $N_1 = N_2 = N/2 - 1$ (red), $N_1=N/2-2,\,N_2=N/2$ (brown), or $N_1=N/2,\,N_2=N/2-2$ (pink). Our analysis shows that, regardless of the user group positioning, the worst effect occurs when users of the shallowest group do not cooperate, implying possible depth effects.

The introduction of reconciliation efficiency has a compressing effect on the rate, making the threshold behavior even more pronounced, as can be seen in Fig.~\ref{threshbplots}(b).

\subsection{\textit{M}-partite systems: \textit{Y}- and \textit{X}-schemes}\label{res3}
\begin{figure}[ht]
    \centering
    \subfloat[\centering]{{
        \includegraphics[width=3.5cm]{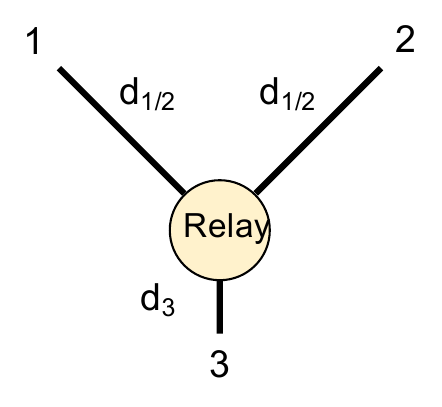} 
        \label{Yschemefig}
    }}%
    \qquad
    \subfloat[\centering]{{
        \includegraphics[width=3.5cm]{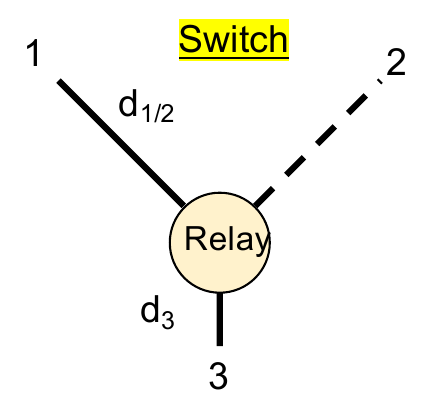} 
        \label{Yschemeswitchfig}
    }}%
	\caption{
 Schematic diagram of a $Y$-scheme with three groups. Group ``$3$" is the deepest in the interferometer, located at a distance of $d_3$ from the relay, while groups $1$" and ``$2$" are equidistant from the relay at a distance of $d_{1/2}$. (b) The relay can also function as a switch, allowing the two superficial groups to communicate directly without passing through the deepest group.
 }
\end{figure}%
We present a study of a specific configuration in which there are $M=3$ groups, each with $N_j=N/M$ users and a pure-loss channel with equal excess noise, $\omega_j=1$ for $j=\left\{1,\,2,\,3\right\}$, resulting in an optimal $M$-partition. The details of this setup are discussed in App.~\ref{tri}. In the $Y$-scheme configuration, the third group, located farthest from the relay, is placed at a distance of $d_3$, while the first and second groups are positioned at an equal distance $d_{1/2}$ from the relay. This configuration is depicted in Fig.~\ref{Yschemefig}. Additionally, the relay can be configured to act as a switch, connecting two groups at a time, as shown in Fig.~\ref{Yschemeswitchfig}.
\begin{figure}[ht]
	\centering
	\includegraphics[width=.9\columnwidth]{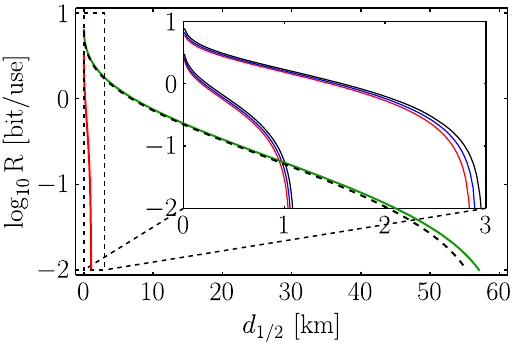}
	\caption{Secret-key rate (in bits per use) as a function of the distance (in kilometers, for standard optical fiber) at which two groups are placed from the relay, while the deepest group in the interferometer is fixed at a distance $d_{deep}$, with $deep=\left\{2,\,3\right\}$. The red, blue, and black curves correspond to distances $d_3$ of $100\,\si{\meter}$, $50\,\si{\meter}$, and $10\,\si{\meter}$, respectively. The secret key rate is robust to changes in the distance $d_3$. Operating the relay as a switch yields a threefold improvement. The inset shows that the performance worsens when going from two to three groups. Specifically, we compare the red curve with $M=2$, $N_j=N/2$ (green) and $M=2$, $N_1=2N/3$, $N_2=N/3$ (black).}
	\label{1gg2}
\end{figure}
\begin{figure}[ht]
	\begin{center}
  		\includegraphics[width=.9\columnwidth]{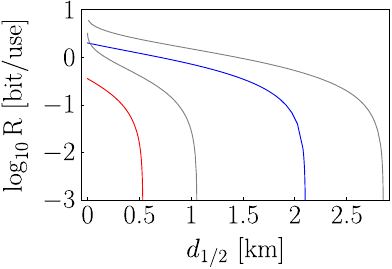}
		\caption{Comparison of the secret-key rate (bit/use) for a non-ideal tripartite $Y$-scheme (red), a switch configuration (blue), and the ideal case (gray), corresponding to Fig.~\ref{1gg2}.}
		\label{NItripY}
	\end{center}
\end{figure}
The impact of the distance $d_{1/2}$ of the shallowest group on the secret-key rate can be seen in Fig.~\ref{1gg2}, where the fixed distance of the deepest group, $d_{3}=\{10\,\si{\meter},\,50\,\si{\meter},\,100\,\si{\meter}\}$, is kept constant. The two sets of curves in the figure represent two different relay configurations, with the right set representing the switch case, which enhances the performance by a factor of nearly three. 

The inset provides a visual comparison to highlight the scaling. When considering an analogous two-group scenario, with the deepest group, i.e., group ``$2$", located $d_2=100\,\si{\meter}$ from the relay, and varying the distance $d_1$ of the other group, secure communication at a distance of nearly $60\,\si{\kilo\meter}$ can be achieved even with a non-optimal split of $N_1=2N/3$ and $N_2=N/3$. However, adding a third group results in a drastic drop in performance that stabilizes immediately afterwards (if other groups are to be added). In a $Y$-scheme with an optimal three-partition, the restriction is to approximately $1\,\si{\kilo\meter}$, while four groups permit secure communication up to approximately $600\,\si{\meter}$. Finally, Fig.~\ref{NItripY} presents the results of the same scenario as Fig.~\ref{1gg2} but with non-ideal Bell detection.
\begin{figure}[ht]
	\centering
	\includegraphics[width=\columnwidth]{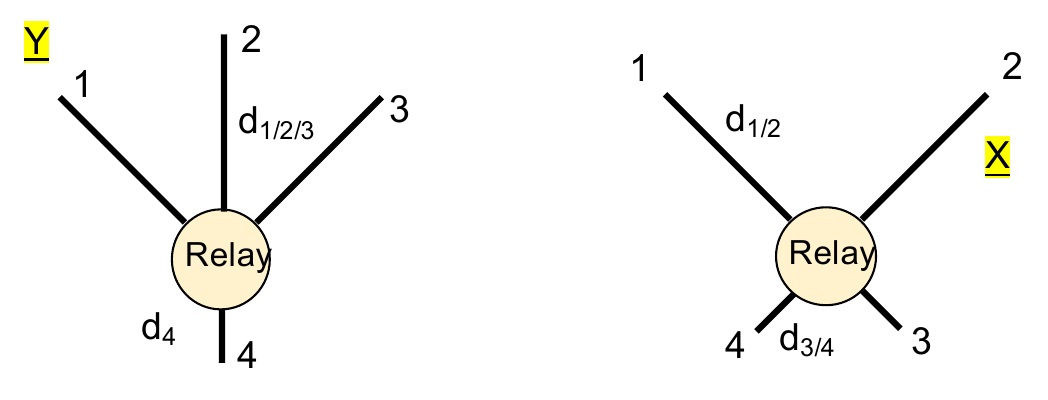}
	\caption{The $Y$-scheme and $X$-scheme are two ways to arrange four groups for secure key distribution.}
	\label{XYscheme4}
\end{figure}%
Fig.~\ref{1gg2} demonstrates the robustness of the secret-key rate to changes in $d_3$ for both operational modes of the relay. The scalability of the scheme is further proven by the extension to four groups in the $Y$-scheme set-up shown in Fig.~\ref{XYscheme4}. The deepest group (group ``$4$") is located at a distance $d_4$ from the relay, while the other three groups (``$1$", ``$2$", and ``$3$") are positioned at an equal distance $d_{1/2/3}$ from the relay. In addition, we analyze another configuration, known as the $X$-scheme, in which the users are distributed with an optimal $M$-partition of the form $N_j=N/M$, where $M=4$. In this configuration, groups ``$1$" and ``$2$" are positioned at a distance $d_{1/2}$ from the relay, while groups ``$3$" and ``$4$" are positioned at a distance $d_{3/4}$ from the relay (as shown in Fig.~\ref{XYscheme4}).
\begin{figure}[ht]
	\centering
	\includegraphics[width=.9\columnwidth]{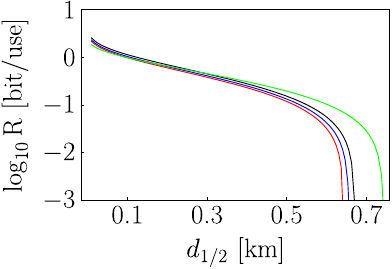}
	\caption{Secret-key rate (bit/use) as a function of the distance (in km, for standard optical fiber) between the relay and the deepest group in the interferometer, for different distances between the other two groups and two fixed values of the deepest group's distance: $d_{deep} = 100\,\si{\meter}$ (red), $50\,\si{\meter}$ (blue), and $10\,\si{\meter}$ (black), corresponding to depths $deep=\left\{4,\,3/4\right\}$, respectively. The $X$-scheme (green) outperforms the $Y$-scheme (red) by less than $15\%$. The robustness to changes in $d_{deep}$ is similar for both schemes.}
	\label{3gg2}
\end{figure}
The secret-key rate of the protocol as a function of the distance $d_{1/2}$ of the first two groups, with the other distances $d_{3/4}$ fixed, is shown in Fig.~\ref{3gg2}. As the best protocol performance is achieved when the deepest group(s) in the interferometer is (are) located closer to the relay, we fix their distance. When distributed in a $X$-scheme (represented by green curves), the four groups perform better (by approximately less than $15\%$) than they would in a $Y$-scheme (represented by red curves). This is because the larger the number of users closer to the relay, the better the performance.

\section{Conclusions}\label{secConcl}
We have presented a novel multipartite CV-MDI-QKD protocol that enables secure quantum secret sharing among an arbitrary number of users. This protocol builds upon the asymmetric configuration from previous works~\cite{PhysRevA.99.030301, Ottaviani2019} and extends the capabilities of standard CV-MDI-QKD~\cite{Pirandola2017, Pirandola2015}. Our analysis focuses on the asymptotic security of the protocol, ignoring finite-size effects and assuming individual uncorrelated attacks. Despite these limitations, the results are promising, especially considering the high level of excess thermal noise we have used in our analysis, which is even higher than what has been achieved experimentally~\cite{Zhang_2019}. Moreover, the challenges associated with modeling a correlated attack make this a highly nontrivial task both theoretically and computationally.

The performance of a $M$-partite CV-MDI-QKD protocol with $M>2$ groups has been analyzed in this study. To simplify the analysis, two specific configurations, the $Y$- and $X$-schemes, have been considered. The results show that the ``switch" variant of the $Y$-scheme leads to improved performance. The protocol also demonstrates robustness to changes in the distance of the deepest group in the interferometer, providing a foundation for building a network of nodes.

In conclusion, it is important to keep in mind that the security of the presented protocol is only proven in the asymptotic limit of many exchanged signals and does not take into account finite-size effects. Further research is needed to improve the security and performance of the protocol. This includes the study of multipartite Bell detections, which have been limited to only a few users so far~\cite{Spedalieri_2013, PhysRevA.93.022325}. Alternatives such as a squeezed state protocol or a thermal-state protocol in the THz frequency range~\footnote{It is attractive for the potential boosting of data rate of wireless communication.}, as well as discrete modulation~\footnote{In order to exploit advantage distillation and post-selection protocols, allowing to improve the achievable distance (paying a price in terms of the key-rate per use of the protocol), one may substitute the Gaussian modulation of the signal states with a discrete one.}~\cite{Pirandola_2020, PhysRevLett.89.167901, PhysRevLett.102.180504}, may lead to improved results. Additionally, exploring other set-ups, such as smaller groups connected with two-by-two Bell-like detections, could help extend the study to more complex networks and clusters of networks. The potential for improvement in this field is vast and provides ample opportunities for future research.

\appendix

\section{Action of the interferometer}\label{appInterf}
The relay station in our model is represented by the $N$-port interferometer outlined in Sec.~\ref{descrip}. This interferometer operates on the travelling modes $\hat{A}$ and is described by the symplectic linear transformation~\cite{Ottaviani2019} given by
\begin{equation}
	\begin{split}
		\hat A_1 \rightarrow A_1&=\frac{1}{\sqrt N}\sum_{j=1}^N\hat A_j, \\
		\hat A_k \rightarrow A_k&=\frac{1}{\sqrt{k\left(k-1\right)}}\left[\left(k-1\right)\,\hat A_k-\sum_{i=1}^{k-1}\hat A_i\right]\\
        \textrm{for}\,\,k&=\left\{2,\,\dots,\,N\right\}.
	\end{split}
\end{equation}% 
For clarity, we will use $A$ instead of $\hat{A}$ to represent the travelling modes after they have undergone the transformation of the interferometer.

\subsection{Bipartite system}
To provide further clarity, let us consider the case where there are only two groups ($M=2$). After the interferometer has acted, the global input state is described by the CM
\begin{equation}\label{esempioVB}
	\textbf{V}_{\textbf B}=
		\left(
		\begin{array}{c|c}
			y\,\boldsymbol I_{2N}	 &\boldsymbol\Upsilon \\
			\hline
			\boldsymbol\Upsilon^T	 &\boldsymbol\Xi  \\	
		\end{array}
		\right),
\end{equation}%
where, for the sake of calculation simplicity, the order of the modes has been changed to $\left\{\hat B_1,\,\dots,\,\hat B_N,\,A_1,\,\dots,\,A_N\right\}$, with $\hat B_j=\left(\hat q_j^B,\,\hat p_j^B\right)^T$ and $A_j=\left(q_j^A,\,p_j^A\right)^T$. Note that in this case, the absence of the hat symbol distinguishes the modes before and after the interferometer's action. In general, for the case $M=2$, the entries of the matrices $\boldsymbol\Upsilon$ and $\boldsymbol\Xi$ can be calculated using Eq.~(\ref{Upsi}). where the value of $\star = \left\{1,\, 2\right\}$ in the expression depends on the group, and with $\Lambda_{a,\,b} := ax_1 + bx_2$. It is noteworthy that, with a proper rearrangement of the modes, the matrix $\boldsymbol\Upsilon$ is upper-triangular. To give a concrete example, let us consider the case where $N=5$, $N_1=2$, and $N_2=3$. This scenario is depicted by the block matrices
\begin{equation}
		\boldsymbol\Upsilon=
		\left(
\begin{array}{ccccc}
\frac{z_1}{\sqrt{5}} & -\frac{z_1}{\sqrt{2}} & -\frac{z_1}{\sqrt{6}} & -\frac{z_1}{2\sqrt{3}} & -\frac{z_1}{2\sqrt{5}} \\
\frac{z_1}{\sqrt{5}} & \frac{z_1}{\sqrt{2}} & -\frac{z_1}{\sqrt{6}} & -\frac{z_1}{2\sqrt{3}} & -\frac{z_1}{2\sqrt{5}} \\
\frac{z_2}{\sqrt{5}} & 0 & \sqrt{\frac{2}{3}}z_2 & -\frac{z_2}{2\sqrt{3}} & -\frac{z_2}{2\sqrt{5}} \\
\frac{z_2}{\sqrt{5}} & 0 & 0 & \frac{\sqrt{3}}{2}z_2 & -\frac{z_2}{2\sqrt{5}} \\
\frac{z_2}{\sqrt{5}} & 0 & 0 &0 & \frac{2z_2}{\sqrt{5}} \\
\end{array}
		\right)\otimes\textbf Z,
\end{equation}%
\begin{equation}		
		\boldsymbol\Xi=
		\left(
\begin{array}{cccccccccc}
\frac{\Lambda_{2,\,3}}{5}& 0 & -\sqrt{\frac{2}{15}}\Lambda_{1,\,-1} & -\frac{\Lambda_{1,\,-1}}{\sqrt{15}} & -\frac{\Lambda_{1,\,-1}}{5} \\
0 & \Lambda_{1,\,0} & 0 & 0 & 0 \\
-\sqrt{\frac{2}{15}}\Lambda_{1,\,-1} & 0 & \frac{\Lambda_{1,\,2}}{3} & \frac{\Lambda_{1,\,-1}}{3\sqrt{2}} & \frac{\Lambda_{1,\,-1}}{\sqrt{30}} \\
-\frac{\Lambda_{1,\,-1}}{\sqrt{15}} & 0 & \frac{\Lambda_{1,\,-1}}{3\sqrt{2}} & \frac{\Lambda_{1,\,5}}{6} & \frac{\Lambda_{1,\,-1}}{2\sqrt{15}} \\
-\frac{\Lambda_{1,\,-1}}{5}& 0 & \frac{\Lambda_{1,\,-1}}{\sqrt{30}} & \frac{\Lambda_{1,\,-1}}{2\sqrt{15}} & \frac{\Lambda_{1,\,9}}{10}\\
\end{array}
		\right)\otimes\textbf I .
\end{equation}
\\
\begin{widetext}
\begin{equation}\label{Upsi}
    \begin{split}
	&\boldsymbol\Upsilon=\left(
		\begin{array}{cccccccc}
			\hlightv{\frac{z_1}{\sqrt N}} & -\frac{z_1}{\sqrt 2} &\cdots&-\frac{z_1}{\sqrt{j\left(j-1\right)}}&\cdots&-\frac{z_1}{\sqrt{k\left(k-1\right)}}&\cdots&-\frac{z_1}{\sqrt{N\left(N-1\right)}}\\
			\vdots & \frac{z_1}{\sqrt 2} &\cdots&\cdots&\cdots&\cdots&\cdots&-\frac{z_1}{\sqrt{N\left(N-1\right)}}\\
			\frac{z_1}{\sqrt N} & 0 & \ddots &\cdots&\cdots&\cdots&\cdots&\vdots\\
			\hline
			\hlighty{\frac{z_2}{\sqrt N}} &0 &\cdots &\sqrt{\frac{j-1}{j}}z_2 &\cdots&\cdots&\cdots&-\frac{z_2}{\sqrt{N\left(N-1\right)}}\\ 
			\vdots &\vdots&\cdots &0 &\ddots&\cdots&\cdots&-\frac{z_2}{\sqrt{N\left(N-1\right)}}\\
			\vdots & \vdots & \cdots &\vdots&0&\sqrt{\frac{k-1}{k}}z_2&\cdots&\vdots\\
			\vdots & 0 & \cdots &\cdots&\cdots&0&\ddots&-\frac{z_2}{\sqrt{N\left(N-1\right)}}\\
			\frac{z_2}{\sqrt N} &0 &\cdots&0&\cdots&\cdots&0&\sqrt{\frac{N-1}{N}}z_2
		\end{array}
		\right)\otimes\textbf{Z},\\
    &\boldsymbol\Upsilon\,\left\{\quad
	\begin{aligned}
		\left\langle\hat B_iA_j\right\rangle&=0,\,\,\,\,\,\,\,\,\qquad\qquad\quad 1\neq i>j, \\
		\left\langle\hat B_mA_1\right\rangle&=\frac{z_\star}{\sqrt N},\,\,\,\,\,\,\qquad\qquad\forall m, \\
		\left\langle\hat B_lA_k\right\rangle&=-\frac{z_\star}{\sqrt{k\left(k-1\right)}},\,\,\,\,\,\quad 1\neq l<k, \\
		\left\langle\hat B_kA_k\right\rangle&=\sqrt{\frac{k-1}{k}}z_\star,\,\,\,\,\,\qquad k\geq 2,
	\end{aligned}\right.,\qquad
    \boldsymbol\Xi\,\left\{\quad
	\begin{aligned}
		\left\langle A_1^2\right\rangle&=\frac{\Lambda_{N_1,\,N_2}}{N}, \\
		\left\langle A_k^2\right\rangle&=\frac{\Lambda_{2,\,k\left(k-1\right)-2}}{k(k-1)},\qquad\qquad\quad k> 2, \\	
		\left\langle A_2A_j\right\rangle&=0,\qquad\qquad\qquad\qquad\quad \forall j\neq 2 \\
		\left\langle A_1A_k\right\rangle&=-2\frac{\Lambda_{1,\,-1}}{\sqrt{Nk\left(k-1\right)}},\,\quad\qquad\, k> 2, \\
		\left\langle A_lA_k\right\rangle&=2\frac{\Lambda_{1,\,-1}}{\sqrt{l\left(l-1\right)k\left(k-1\right)}},\,\,\,\,\quad l,\,k> 2,
	\end{aligned}\right.
    \end{split}
\end{equation}
\end{widetext}

\section{Generalised multipartite Bell detection}\label{bell}
To perform the multipartite Bell detection, $N-1$ homodyne detections in the $\hat q$-quadrature and one homodyne detection in the $\hat p$-quadrature are carried out. Using the example in Eq.~(\ref{esempioVB}), in the scenario with $N=5$, $N_1=2$, and $N_2=3$, the resulting conditional global input state is described as
\begin{equation}\label{N5c}
	\textbf{V}_{\textbf B|\gamma}=\left(
		\begin{array}{ccccc}
			\tikzmark{left}{\textbf A}&\textbf C&\textbf E&\textbf E&\textbf E\\
			\textbf C&\tikzmark{right}{\textbf A}\Hlightg[first]&\textbf E&\textbf E&\textbf E\\
			\textbf E&\textbf E&\tikzmark{left}{\textbf B}&\textbf D&\textbf D\\
			\textbf E&\textbf E&\textbf D&\textbf B&\textbf D\\
			\textbf E&\textbf E&\textbf D&\textbf D&\tikzmark{right}{\textbf B}\\
		\end{array}\right).\Hlightr[first]
\end{equation}%
where
\begin{equation}
	\begin{split}
	\textbf{A}&=y\textbf I-\left(
		\begin{array}{cc}
 			\frac{1}{x_1}\frac{\Lambda_{3,\,1}}{\Lambda_{3,\,2}} & 0\\
			0 & \frac{1}{\Lambda_{2,\,3}}
		\end{array}\right)z_1^2,\\
	\textbf{B}&=y\textbf I-\left(
		\begin{array}{cc}
 			\frac{2}{x_2}\frac{\Lambda_{1,\,1}}{\Lambda_{3,\,2}} & 0\\
			0 & \frac{1}{\Lambda_{2,\,3}}
		\end{array}\right)z_2^2,\\
	\textbf{C}&=\left(
		\begin{array}{cc}
 			\frac{x_2 }{x_1}\frac{1}{\Lambda_{3,\,2}} & 0\\
			0 & -\frac{1}{\Lambda_{2,\,3}}
		\end{array}\right)z_1^2,\\
	\textbf{D}&=\left(
		\begin{array}{cc}
 			\frac{x_1 }{x_2}\frac{1}{\Lambda_{3,\,2}} & 0\\
			0 & -\frac{1}{\Lambda_{2,\,3}}
		\end{array}\right)z_2^2,\\	
	\textbf{E}&=\left(
		\begin{array}{cc}
 			\frac{1}{\Lambda_{3,\,2}} & 0\\
			0 & -\frac{1}{\Lambda_{2,\,3}}\\
		\end{array}\right)z_1 z_2.		
	\end{split}
\end{equation}%

\section{Unitary entanglement localisation of \textit{M}-symmetric states}\label{reduction}
Eq.~(\ref{N5c}) displays a distinctive symmetry that remains consistent for any value of $M$, making it possible to simplify our problem. To demonstrate this simplification, let us examine one quadrature (the same reasoning can be applied to the other). As a straightforward application of linear algebra~\cite{PhysRevA.71.032349}, let us consider a $N\times N$ matrix of the form
\begin{equation}\label{WNreduction}
	\begin{split}
		\textbf{W}_{N_j}:=&\,\left(d_j-c_j\right)\,\textbf I_{N_j}+N_jc_j\,\textbf P_{N_j}\\	
		=&\left(
		\begin{array}{ccccc}
			d_j	 &c_j &c_j &\cdots &c_j \\
			c_j	 &d_j &c_j &\ddots &c_j \\
			c_j	 &c_j &d_j &\ddots &c_j \\
			\vdots	 &\ddots &\ddots &\ddots &\vdots \\
			c_j	 &c_j &c_j &c_j &d_j
		\end{array}
		\right),
	\end{split}
\end{equation}%
where $\textbf P_{N_j}$ denotes the projection matrix onto the vector $v_{N_j}=N_j^{-1/2}\left(1,\,1,\,\dots,\,1\right)^T$~\footnote{It is the matrix with all elements equal to $N_j^{-1}$.}. With the above Eq.~(\ref{WNreduction}), it is straightforward to see that the matrix is diagonal in the basis defined by $v_{N_j}$ and $N_j-1$ orthogonal vectors, that is
\begin{equation}\label{WpNireduction}
	\begin{split}
		\textbf W'_{N_j}&=\textbf R_{N_j}^{-1}\textbf W_{N_j}\textbf R_{N_j}\\
		&=\left(
		\begin{array}{ccccc}
			d_j-c_j	 &0 &0 &\cdots &0 \\
			0	 &d_j-c_j &0 &\ddots &0 \\
			0	 &0 &d_j-c_j &\ddots &0 \\
			\vdots	 &\ddots &\ddots &\ddots &\vdots \\
			0	 &0 &0 &0 &d_j+\left(N_j-1\right)c_j
		\end{array}
		\right).
	\end{split}		
\end{equation}%
The matrix $\textbf R_{N_j}$ is the rotation that diagonalizes the matrix, which can be obtained from the basis of eigenvectors $\left\{e_k\right\}_{k=1}^{N_j}$ of the matrix itself. It is given by $\textbf R_{N_j}=N_j^{-1/2}\left(e_1,\,\dots,\,e_{N_j}\right)^T$. We define
\begin{defn}
    An $M$-symmetric state is a multi-partite state of $\sum_{j=1}^M N_j$ modes characterized by its CM $\textbf V_{\textbf M}$. The state is constructed by incorporating diagonal blocks,
    \begin{equation}
		\textbf O_{N_jN_j}=\left(d_j-c_j\right)\,\textbf I_{N_j}+N_jc_j\,\textbf P_{N_j}\equiv\textbf{W}_{N_j},
	\end{equation}%
    with the same symmetry as $\textbf {W}_{N_j}$, and off-diagonal blocks,
    \begin{equation}
		\textbf O_{N_iN_j}\equiv\textbf P_{f_{ij}^{-1}}\quad\left(i\neq j\right),
	\end{equation}%
    which are proportional to $\textbf P_{N_j}$ and have all elements equal to $f_{ij}$.
\end{defn}%
For clarity, we present an example of a
\begin{equation}
	\textbf V_{\textbf 3}=\left(
	\begin{array}{ccc}
		\textbf{W}_{N_1}& \textbf P_{f_{12}^{-1}}&\textbf P_{f_{13}^{-1}}\\
		\textbf P_{f_{12}^{-1}}& \textbf{W}_{N_2}&\textbf P_{f_{23}^{-1}}\\
		\textbf P_{f_{13}^{-1}}& \textbf P_{f_{23}^{-1}}&\textbf{W}_{N_3}\\
	\end{array}
		\right).
\end{equation}%
As previously stated, Eq.~(\ref{N5c}) is an example of a particular $\textbf V_{\textbf 2}$. By applying the same reasoning as in Eq.~(\ref{WpNireduction}), we can find the transformed CM of a general $M$-symmetric state, as
\begin{equation}
    \textbf V'_{\textbf M}=\bigoplus_{i=1}^M\textbf R_{N_i}^{-1}\textbf V_{\textbf M}\bigoplus_{j=1}^M\textbf R_{N_j},
\end{equation}%
whose blocks are therefore simply given by
\begin{equation}
	\textbf O'_{N_iN_j}=\textbf R_{N_i}^{-1}\textbf O_{N_iN_j}\textbf R_{N_j}.
\end{equation}%
Thus, the transformed CM $\textbf V'{\textbf M}$ describes an effective state of $M$ modes, since $\left(\sum_{j=1}^M N_j\right)-M$ of them are thermal (or vacuum) states that are uncorrelated with each other [as seen in Eq.~(\ref{eqexvpd})]. The effective $M$-mode state is described by
\begin{equation}
	\textbf O'_{N_iN_j}=\left[d_i+\left(N_i-1\right)\,c_i\right]\delta_{ij}+\left(1-\delta_{ij}\right)\,f_{ij}\sqrt{N_iN_j}.
\end{equation}%
For example, we provide a transformed
\begin{widetext}
\begin{equation}\label{eqexvpd}
\textbf{V}'_{\textbf 2}=
	\begin{tikzpicture}[baseline={(0,0)}]
		\matrix [matrix of math nodes,left delimiter=(,right delimiter=)] (m)
       		{
		d_1-c_1 &0 &\ddots &0 &0&0&0&0 \\
		0 &d_1-c_1 &\ddots &0 &0&0&0&0 \\
		\vdots &\ddots &\ddots &\vdots &0&0&0&0 \\
		0 &0 &0 &d_1+\left(N_1-1\right)c_1 &f_{12}\sqrt{N_1N_2}&0&0&0 \\
		0&0&0&f_{12}\sqrt{N_1N_2}&d_2+\left(N_2-1\right)c_2 &0&0&0\\
		0&0&0&0&\vdots&\ddots&\ddots&\vdots\\
		0&0&0&0&0&\ddots&d_2-c_2&0\\
		0&0&0&0&0&\ddots&0&d_2-c_2\\
      		};
      		\draw[rounded corners, thick, draw, fill=blue!50, fill opacity=0.2,text opacity =1,inner sep=0pt] (m-4-4.north west) rectangle (m-5-5.south east);             
	\end{tikzpicture}\,,
\end{equation}	
\end{widetext}%
and therefore $\textbf V'_{\textbf 2}$ corresponds $ N_1+N_2-2$ uncorrelated thermal modes, while the effective $2$-mode state is described by
\begin{equation}
	\textbf{V}'_{2}=
	\begin{tikzpicture}[baseline={(0,0)}]
		\matrix [matrix of math nodes,left delimiter=(,right delimiter=)] (m)
     		{
		d_1+\left(N_1-1\right)c_1 &f_{12}\sqrt{N_1N_2} \\
		f_{12}\sqrt{N_1N_2} &d_2+\left(N_2-1\right)c_2 \\
       		};
        		\draw[rounded corners, thick, draw, fill=blue!50, fill opacity=0.2,text opacity =1,inner sep=0pt] (m-1-1.north west) rectangle (m-2-2.south east);             
	\end{tikzpicture}\,.
\end{equation}%    
By induction and following the above reasoning, the general post-reduction CM of the effective $M$-mode state is given by Eqs.~\ref{VMgamma} and~\ref{gammaij} in the main text. To clarify these expressions, let us consider the case of
\begin{equation}
	\textbf V_{3|\gamma}=\left(
		\begin{array}{ccc}
			\boldsymbol \Gamma_{11}&\boldsymbol \Gamma_{12}&\boldsymbol \Gamma_{13}\\
			\boldsymbol \Gamma_{12}&\boldsymbol \Gamma_{22}&\boldsymbol \Gamma_{23}\\
			\boldsymbol \Gamma_{13}&\boldsymbol \Gamma_{23}&\boldsymbol \Gamma_{33}\\
		\end{array}\right).
\end{equation}%
where
\begin{equation}\label{eqexred2}
	\begin{split}
	\boldsymbol \Gamma_{11}&=y\textbf I-\left(
		\begin{array}{cc}
 			\frac{N_3x_2+N_2x_3}{\Theta_q} & 0\\
			0& \frac{N_1}{\Theta_p}
		\end{array}\right)z_1^2,\\
	\boldsymbol \Gamma_{22}&=y\textbf I-\left(
		\begin{array}{cc}
 			\frac{N_3x_1+N_1x_3}{\Theta_q} & 0\\
			0&\frac{N_2}{\Theta_p}
		\end{array}\right)z_2^2,\\
	\boldsymbol \Gamma_{33}&=y\textbf I-\left(
		\begin{array}{cc}
 			\frac{\Lambda_{N_2,\,N_1}}{\Theta_q} & 0\\
			0& \frac{N_3}{\Theta_p}
		\end{array}\right)z_3^2,\\
	\boldsymbol \Gamma_{12}&=\sqrt{N_1N_2}\left(
		\begin{array}{cc}
 			\frac{x_3}{\Theta_q} & 0\\
			0 & -\frac{1}{\Theta_p}
		\end{array}\right)z_1z_2,\\	
	\boldsymbol \Gamma_{13}&=\sqrt{N_1N_3}\left(
		\begin{array}{cc}
 			\frac{x_2}{\Theta_q} & 0\\
			0 & -\frac{1}{\Theta_p}
		\end{array}\right)z_1z_3,\\
	\boldsymbol \Gamma_{23}&=\sqrt{N_2N_3}\left(
		\begin{array}{cc}
 			 \frac{x_1}{\Theta_q} & 0\\
			0 &-\frac{1}{\Theta_p}
		\end{array}\right)z_2z_3,
	\end{split}
\end{equation}%
with $\Theta_q:=N_1x_2x_3+cyclics$ and $\Theta_p:=\sum_{j=1}^3N_jx_j$. The change in labeling is made to facilitate comprehension [see Eqs.~\ref{eqexred2} and~\ref{gammaij}].

\section{Secret-key rate}\label{appRate}
Before the action of the eavesdropper and the measurements, the global input state that describes the parties (the Bobs) and the eavesdropper (Eve) is pure and Gaussian. After her action and before the measurements, the global output state is still pure, although it may be non-Gaussian. The local measurements commute, so we can defer the Bobs' heterodyne detections until after Eve's measurement. As a result, the Bobs and Eve share a pure conditional state $\hat\rho_{\textbf BE|\gamma}$, where we label the local modes as $\hat{\textbf B}:= \hat B_1\cdots \hat B_N$. The reduced states for the Bobs and Eve are $\hat\rho_{\textbf B|\gamma}$ and $\hat\rho_{\textbf BE|\gamma}$, respectively. Since the conditional state is pure, the von Neumann entropies $S$ of the subsystems are equal, meaning
\begin{equation}
	S\left(\hat\rho_{\textbf B|\gamma}\right)=S\left(\hat\rho_{E|\gamma}\right).
\end{equation}%
Analogously, in the conditional post-relay scheme, the action of the Bobs projects $\hat\rho_{\textbf BE|\gamma}$ into a pure~\footnote{This occurs because heterodyne detection is a rank-$1$ measurement, hence the purity.} state $\hat\rho_{\textbf BE|\gamma\tilde\beta^{(N)}}$, yielding to
\begin{equation}
	S\left(\hat\rho_{\textbf B|\gamma\tilde\beta^{(N)}}\right)=S\left(\hat\rho_{E|\gamma\tilde\beta^{(N)}}\right).
\end{equation}%
As a consequence, the amount of information that Eve can obtain about Bobs' variables $\tilde\beta^{(N)}:=\left\{\tilde\beta_j\right\}_{j=1}^N$, conditioned on $\gamma$, is upper-bounded by her Holevo quantity
\begin{equation}
    \begin{split}
        \chi_{E|\gamma}&=S\left(\hat\rho_{E|\gamma}\right)-S\left(\hat\rho_{E|\gamma\tilde\beta^{(N)}}\right)\\
        &=S\left(\hat\rho_{\textbf B|\gamma}\right)-S\left(\hat\rho_{\textbf B|\gamma\tilde\beta^{(N)}}\right),
    \end{split}
\end{equation}%
which is fully determined by the conditional state $\hat\rho_{\textbf B|\gamma}$. Indeed, assuming asymptotic security and infinite Gaussian modulation, the secret-key rate of the protocol can then be expressed as
\begin{equation}
	R=\xi I_{\textbf B|\gamma}-\chi_{E|\gamma},
\end{equation}%
where $\xi<1$ is the reconciliation efficiency. It is worth noting that, even though $\sum_{j=1}^M N_j < N$, Eve still has a purification of the global state due to the assumption of all trusted users, and as a result, the secret-key rate is covariant~\footnote{In the sense that preserves Eq.~(\ref{erreibgammachiegamma}) in form.}~\cite{Ottaviani2019}.

\section{Bipartite system}\label{bipartsys}
In the QSS protocol, the parties are divided into $M=2$ groups, and the effective two-mode CM can be obtained following the methods presented in Refs.~\cite{Ottaviani2019}. Using the same notation, the resulting CM is given by
\begin{equation}\label{Vabg}
		\textbf V_{2|\gamma}=\left(
		\begin{array}{cc}
			\boldsymbol \Delta_1 & \boldsymbol \Gamma'	\\
			\boldsymbol \Gamma' & \boldsymbol \Delta_2 	\\ 	
		\end{array}
		\right),
\end{equation}%
where, for $l=\{1,\,2\}$, one explicitly has
\begin{equation}\label{elem}
	\begin{split}
		\boldsymbol \Delta_l&=y-\textrm{diag}\left(\frac{\left(N-N_l\right) z_l^2}{\Lambda_{N-N_1,\,N-N_2}},\, \frac{N_l z_l^2}{\Lambda_{N-N_2,\,N-N_1}}\right), \\
		\boldsymbol \Gamma'&=z_1z_2\sqrt{N_1N_2}\,\textrm{diag}\left(\frac{1}{\Lambda_{N-N_1,\,N-N_2}},\,-\frac{1}{\Lambda_{N-N_2,\,N-N_1}}\right),
	\end{split}
\end{equation}%
and again $\Lambda_{a,\,b}:= ax_1+bx_2$. One may compute the symplectic eigenvalues of the CM Eq.~(\ref{Vabg}) as~\cite{RevModPhys.84.621} 
\begin{equation}
	\nu_\pm = \sqrt{\frac{\Delta\pm\sqrt{\Delta^2-4\left\|\textbf V_{2|\gamma}\right\|}}{2}},
\end{equation}%
where $\Delta=\left\|\boldsymbol\Delta_1\right\|+\left\|\boldsymbol\Delta_2\right\|+2\left\|\boldsymbol\Gamma'\right\|$ and $\left\|\cdot\right\|$ indicates the determinant. However, it is not known beforehand if the rate will be asymptotically maximum or if there exists an optimal modulation value $\mu$ that maximizes it. Our analysis shows that there is an optimal modulation in the full-house case, where all users participate. Additionally, we also study the asymptotic trends for large modulation values in the full-house (FH) scenario and find that
\begin{equation}\label{nupm}
	\begin{split}
		\nu_+^{(FH)}&\to\left|\eta_1-\eta_2\right|\sqrt{\frac{N_1N_2}{\mathcal N_{12}\widetilde{\mathcal N}_{12}}}\,\mu,\\
		\nu_-^{(FH)}&\to\frac{1}{\left|\eta_1-\eta_2\right|}\sqrt{\frac{\lambda_{12}\widetilde\lambda_{12}}{N_1N_2}},
	\end{split}	
\end{equation}%
with
\begin{equation}
	\begin{split}
		\lambda_{ij}&:= N_i \omega_j\left(1 - \eta_j\right)+N_j \omega_i\left(1 - \eta_i\right),\\
		\widetilde\lambda_{ij}&:= N_i \omega_i\left(1 - \eta_i\right)+N_j \omega_j\left(1 - \eta_j\right),\\
		\mathcal N_{ij}&:= N_i\eta_j+N_j\eta_i,\\
		\widetilde{\mathcal N}_{ij}&:= N_i\eta_i+N_j\eta_j,
	\end{split}
\end{equation}%
which are all symmetrical with respect to the interchange of indices $i$ and $j$. The removable discontinuity $\eta_1=\eta_2$ does not represent a problem, as we will prove in the next Sec.~\ref{appcondhetdet}.

\subsection{Conditioning: Heterodyne Detection}\label{appcondhetdet}
Assuming group ``$2$" serves as the decoder, the conditional CM after local heterodyne detection of Eq.~(\ref{Vabg}) is calcalated as~\cite{RevModPhys.84.621}
\begin{equation}\label{Vagb}
	\textbf V_{1|\gamma 2}=\boldsymbol \Delta_1-\boldsymbol \Gamma'\left(\textbf I + \boldsymbol \Delta_2\right)^{-1}\boldsymbol \Gamma'^{\textrm{T}},
\end{equation}%
which is diagonal and therefore its symplectic eigenvalue can be obtained by the symplectic invariance of its determinant, that is
\begin{equation}\label{nunss}
	\nu_N^{SS}=\sqrt{\left\|\textbf V_{1|\gamma 2}\right\|}.
\end{equation}%
Specifically, in the FH limit, the symplectic eigenvalue is given by
\begin{equation}\label{nuNSS}
	\nu_N^{SS}\rightarrow\frac{1}{\eta_1}\sqrt{\frac{\left(\lambda_{12}+N_1\eta_2\right)\left(\widetilde\lambda_{12}+N_2\eta_2\right)}{N_1N_2}}.
\end{equation}%
Having the total and conditional symplectic spectra [Eqs.~(\ref{nupm}) and~(\ref{nuNSS}), respectively], the Holevo quantity using  as
\begin{equation}
	\chi=h\left(\nu_+\right)+h\left(\nu_-\right)-h\left(\nu_N^{SS}\right),
\end{equation}%
where the \textit{entropic function} $h$ is defined as
\begin{equation}
		h(\nu):=\frac{\nu+1}{2}\log_2\left(\frac{\nu+1}{2}\right)-\frac{\nu-1}{2}\log_2\left(\frac{\nu-1}{2}\right).
\end{equation}%
The function is equal to zero for the vacuum noise $h\left(1\right)=0$ and asymptotically approaches
\begin{equation}
		h(\nu)=\log_2{\frac{\ee}{2}\nu}+O\left(\nu^{-1}\right).
\end{equation}%
The continuity of the Holevo quantity $\chi$ in the transition from the asymmetrical to the symmetrical configuration, i.e., in $\left\{\eta_1=\eta_2,\,\omega_1=\omega_2\right\}$, must be verified. This can be done by comparing their respective symplectic eigenvalues in the FH case. When $\eta_1=\eta_2:=\eta$ and $\omega_1=\omega_2:=\omega$, the following holds
\begin{equation}\label{nupmcont}
 	\begin{split}
		\nu_\pm^{(FH)}&\to\sqrt{y\left(y-\frac{z^2}{x}\right)}\\
		&=\sqrt{\frac{\left[\left(1-\eta\right)\mu\omega+\eta\right]\mu}{\left(1-\eta\right)\omega+\eta\mu}},
	\end{split}
\end{equation}%
in agreement with Refs.~\cite{Ottaviani2019}. The same continuity holds for $\nu_N^{SS}$, which converges to
\begin{equation}
	\nu_N^{SS}\to\sqrt{\frac{\tau_{12}\tau_{21}}{\widetilde\tau_{12}\widetilde\tau_{21}}},
\end{equation}%
where we define 
\begin{equation}
	\begin{aligned}
		\tau_{ij}&:= \eta\left(N_i+N_j\mu\right)+N\omega\left(1-\eta\right)\mu \\
		\widetilde\tau_{ij}&:=\eta\left(N_i+N_j\eta\mu\right)+N\omega\left(1-\eta\right). \\
	\end{aligned}
\end{equation}%

\subsection{Mutual Information}
The mutual information can be expressed compactly as~\cite{Pirandola2015}
\begin{equation}\label{mutuainfoconSigma}
	I=\frac{1}{2}\log_2\Sigma,
\end{equation}%
where we make the assumption that group ``$2$" serves as the decoder, allowing us to write
\begin{equation}
	\Sigma=\frac{1+\left\|\boldsymbol\Delta_1\right\|+\textrm{Tr}\left\{\boldsymbol\Delta_1\right\}}{1+\left\|\textbf V_{1|\gamma 2}\right\|+\textrm{Tr}\left\{\textbf V_{1|\gamma 2}\right\}},
\end{equation}%
with $\boldsymbol\Delta_1$ and $\textbf V_{1|\gamma 2}$ defined in Eqs.~(\ref{Vabg}) and~(\ref{Vagb}), respectively. A closer examination of the denominator $\sigma_n:=1+\left\|\textbf V_{1|\gamma 2}\right\|+\textrm{Tr}\left\{\textbf V_{1|\gamma 2}\right\}$ reveals
\begin{equation}
    \begin{split}
        &\sigma_n\to\\
        &\frac{\eta_2\left(\eta_1+\eta_2\right)N\left(N-N_1-N_2\right)^2\mu^2}{\left(N-N_1\right)\left[N\left(\eta_1+\eta_2\right)-\mathcal N_{12}\right]\left[\left(N-N_2\right)\left(\eta_1+\eta_2\right)-N_1\eta_2\right]}.
    \end{split}
\end{equation}%
The (quadratic) dependence on the modulation $\mu$ highlights the importance of identifying an optimal value of $\mu$ to maximize the secure communication performance. In contrast, in the FH case, there is no such dependence, as
\begin{equation}
		\sigma_n^{(FH)}\to\frac{\left(\lambda_{12}+\mathcal N_{12}\right)\left(\widetilde\lambda_{12}+\widetilde{\mathcal N}_{12}\right)}{N_1N_2\eta_1^2}.
\end{equation}%
This suggests that for a bipartite system, there is a direct relationship between full-house and asymptotic behavior. Whenever all users cooperate, the rate is maximized for high values of modulation $\mu\gg1$. On the other hand, as soon as one or more users do not cooperate, there exists an optimal modulation value that maximizes the rate.

\subsection{Secret-key rate}
In a thermal-loss channel with asymptotic security, the secret-key rate against collective attacks can be obtained using Eq.~(\ref{erreibgammachiegamma}), assuming perfect reconciliation ($\xi=1$) and utilizing an infinite Gaussian modulation as
\begin{equation}\label{Rnormale}
	R=\frac{1}{2}\log_2\Sigma-h\left(\nu_+\right)-h\left(\nu_-\right)+h\left(\nu_N^{SS}\right).
\end{equation}%
One may determine its FH asymptotic limit, resulting in 
\begin{equation}\label{Rasy}
	\begin{split}
		R^{asy}&=\log_2\left(\frac{2\eta_1\eta_2}{\ee\left|\eta_1-\eta_2\right|}\sqrt{\frac{N_1N_2}{\left(\lambda_{12}  + \mathcal N_{12}\right)\left(\widetilde\lambda_{12} + \widetilde{\mathcal N}_{12}\right)}}\right)\\
		&-h\left(\frac{1}{\left|\eta_1 - \eta_2\right|}\sqrt{\frac{\lambda_{12}\widetilde\lambda_{12}}{N_1 N_2}}\,\right)\\
		&+h\left[\frac{1}{\eta_1}\sqrt{\frac{\left(\lambda_{12} + N_1 \eta_2\right)\left(\widetilde\lambda_{12}+ N_2 \eta_2\right)}{N_1 N_2}}\right].
	\end{split}
\end{equation}%
\begin{widetext}
The asymmetric configuration, when ideal conditions are met, enables secure long-distance communication. In particular, for $\eta_2=1$ (which corresponds to a distance of $0,\si{\kilo\meter}$ for the second user), the secret-key rate expression in Eq.~(\ref{Rasy}) simplifies to

\begin{equation}
	\begin{split}
		R^{asy}\left(\eta_2=1\right)&=\log_2\left(\frac{2\eta_1}{e\left(1-\eta_1\right)}\sqrt{\frac{N_1N_2}{\left\{N_1+N_2\left[\omega_1\left(1-\eta_1\right)+\eta_1\right]\right\}\left\{N_2+N_1\left[\omega_1\left(1-\eta_1\right)+\eta_1\right]\right\}}}\right)\\
		&-h\left(\omega_1\right)+h\left\{\frac{1}{\eta_1}\sqrt{\frac{\left[N_1+N_2\omega_1\left(1-\eta_1\right)\right]\left[N_2+N_1\omega_1\left(1-\eta_1\right)\right]}{N_1 N_2}}\right\}.
	\end{split}
\end{equation}
\end{widetext}%
Under the condition that group ``$1$" has pure-loss links ($\omega_1=1$), Eq.~(\ref{Rasy}) can be further simplified to
\begin{equation}
	\begin{split}
		R^{asy}&\left(\eta_2=1,\,\omega_1=1\right)=\log_2\left[\frac{\eta_1}{\ee\left(1-\eta_1\right)}\frac{\sqrt{N_1N_2}}{N}\right]\\
		&+h\left\{\frac{1}{\eta_1}\sqrt{\frac{\left[N_1+N_2\left(1-\eta_1\right)\right]\left[N_2+N_1\left(1-\eta_1\right)\right]}{N_1 N_2}}\right\}.
	\end{split}
\end{equation}
\begin{figure}[ht]
	\begin{center}
  		\includegraphics[width=.9\columnwidth]{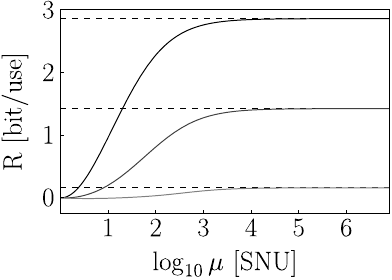}
		\caption{The solid curves show the secret-key rate (bit/use) as a function of modulation $\mu$, obtained from Eq.~(\ref{Rnormale}), for different splittings of the signal in the full-house configuration, where Alice sends signals to all four users. The dotted curves represent the asymptotic rate Eq.~(\ref{Rasy}) for large values of $\mu$. Moving from top to bottom, the splitting ratios are $50/50$, $5/95$, and $1/99$1. The distances and thermal noises are fixed to $d_1=1\,\si{\kilo\meter}$, $d_2=0.1\,\si{\kilo\meter}$, and $\omega_1=\omega_2=1$ SNU. The figure shows that the optimal modulation for maximizing the rate is always large, independent of the splitting, and that the performance is symmetric with respect to the `X/Y" and ``Y/X" splittings.}
		\label{approach}
	\end{center}
\end{figure}%
\noindent The rate Eq.~(\ref{Rasy}) is based on the assumption of infinite use of the relay channel. However, this can be closely approximated after a large but finite number of rounds, as demonstrated in Fig.~\ref{approach}. The fast convergence of the rate expressed in Eq.~(\ref{Rnormale}) to its asymptotic value is particularly noteworthy. Additionally, it is observed that the configurations ``X/Y" and ``Y/X" show the same behavior, indicating that there are no depth effects introduced by the relay. Furthermore, when $N_1 + N_2 < N$, there exists an optimal modulation value $\mu$ that maximizes the rate, as confirmed by Fig.~\ref{0gg} in the case of a pure-loss channel with $\omega_1 = \omega_2 = 1$.
\begin{figure}[ht]
		\centering
		\includegraphics[width=.9\columnwidth]{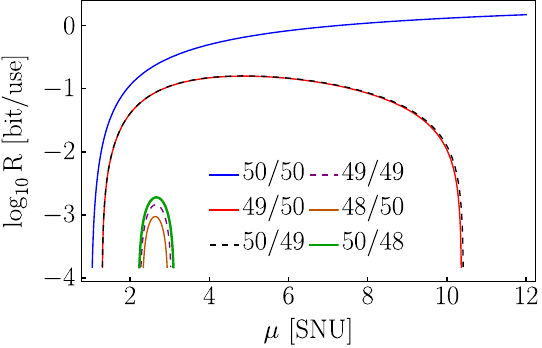}
	\caption{The secret-key rate (bit/use) as a function of the modulation $\mu$ (SNU) for a pure-loss channel (i.e., $\omega_1=\omega_2=1$) exhibits the presence of an optimal $\mu$ for dummy users, regardless of the distance of the groups from the relay. The curves show the performance of different ``asymmetries" in the $50/50$ splitting. The blue curve corresponds to the optimal $50/50$ case, which has the same trend shown in Fig.~\ref{approach}. Other parameters are $d_1=1$ km and $d_2=0.01$ km.}
	\label{0gg}
\end{figure}

\subsection{Non-ideal Bell detector}	
In the bipartite asymmetrical scenario, the transformation rule Eq.~(\ref{elem}) is represented as
\begin{equation}\label{xniasim}
	\begin{split}
		\left(N-N_1\right)x_1&+\left(N-N_2\right)x_2\\
		&\mapsto \left(N-N_1\right)x_1+\left(N-N_2\right)x_2+N\frac{1-\tau}{\tau}, \\
		\left(N-N_2\right)x_1&+\left(N-N_1\right)x_2\\
		&\mapsto \left(N-N_2\right)x_1+\left(N-N_1\right)x_2+N\frac{1-\tau}{\tau}.
	\end{split}	
\end{equation}
This generalizes the results from Refs.~\cite{PhysRevA.99.030301, Spedalieri_2013, Vanmeter2014}. Furthermore, in the case of a FH scenario, with $N_1+N_2=N$, Eq.~(\ref{xniasim}) simplifies to
\begin{equation}\label{xniasimFH}
		x_j\mapsto x_j+\frac{1-\tau}{\tau}, \qquad j=\{1,\,2\}.
\end{equation}%
One can then generalize Eq.~(\ref{Rasy}) by following the approach presented in Ref.~\cite{Spedalieri_2013} and applying the transformations from Eq.~(\ref{xniasimFH}). Although Eq.~(\ref{Rasy}) is not expressed in terms of $x_j$, $y$, and $z_j$, but rather in terms of the channel parameters $\eta_j$, $\omega_j$, and the modulation $\mu$, making this substitution non-trivial, after developing the usual analysis, it can be shown that, in the FH case, the asymptotic non-ideal secret-key rate for two groups with a thermal-loss channel is given by
\begin{equation}
	\begin{split}
		R^{asy}&=\log_2\left(\frac{2\tau\eta_1\eta_2}{\ee\left|\eta_1-\eta_2\right|}\sqrt{\frac{N_1N_2}{L_{12}L_{21}}}\right)-h\left(\frac{1}{\tau\left|\eta_1 - \eta_2\right|}\sqrt{\frac{S_{12} S_{21}}{N_1 N_2}}\right)\\
		&+h\left(\frac{1}{\tau\eta_1}\sqrt{\frac{R_{12}R_{21}}{N_1 N_2}}\right),
	\end{split}
\end{equation}%
where 
\begin{equation}
	\begin{split}
		S_{ij}&=N_i\left[1-\tau+\tau\omega_1\left(1-\eta_1\right)\right]\\
		&+N_j\left[1-\tau+\tau\omega_2\left(1-\eta_2\right)\right],\\
		R_{ij}&=N_i\left[1-\tau+\tau\omega_1\left(1-\eta_1\right)\right]\\
		&+N_j\left[1-\tau\left(1-\eta_2\right)+\tau\omega_2\left(1-\eta_2\right)\right],\\
		L_{ij}&=N_i\left[1-\tau\left(1-\eta_1\right)+\tau\omega_1\left(1-\eta_1\right)\right]\\
		&+N_j\left[1-\tau\left(1-\eta_2\right)+\tau\omega_2\left(1-\eta_2\right)\right].
	\end{split}
\end{equation}%

\section{\textit{M}-partite systems: \textit{Y}- and \textit{X}-schemes}\label{tri}
The standard procedure is followed when analyzing the security of a system with more than two groups. However, when dealing with multiple groups, it is important to pay attention to the rate. The smallest possible secret-key rate is obtained by subtracting the maximum Holevo quantity $\chi$ from the minimum mutual information $I$ between any two groups. This is because the rate between different groups can vary and a potential eavesdropper may attack the group with the lower rate. To consider the worst-case scenario, we need to take the lowest possible rate into account.

To optimize the system, we must first find the symplectic eigenspectrum of $\textbf V_{M|\gamma}$ [see Eq.~(\ref{VMgamma})]. The conditioning of the system can then be performed in $M$ different ways and we need to find the method that maximizes the Holevo quantity or minimizes the conditional von Neumann entropy $S_{cond}$. To do this, we perform local heterodyne detection on $\textbf V_{M|\gamma}$ and find the group that is furthest from the relay, as this group will result in the minimum conditional entropy~\footnote{This behaviour does not depend on the modulation $\mu$. In case there are more equidistant groups further away from the relay, it makes no difference which group is considered.}.

For the switch to function correctly, we must consider all combinations of two groups and perform a heterodyne measurement. They are characterised by the $\binom{M}{2}$ $\textbf V_{yz}s$ $4\times 4$ matrices built from $\textbf V_{M|\gamma}$ with the blocks $\boldsymbol\Gamma_{yy}$, $\boldsymbol\Gamma_{zz}$, and $\boldsymbol\Gamma_{yz}$ of the groups ``Y" and ``Z" of interest, with $y$, $z=\left\{1,\,\dots,\,M\right\}$. The minimum is still obtained by measuring the group that is furthest from the relay.

The correct mutual information Eq.~(\ref{mutuainfoconSigma}) can be determined by considering all the $M\left(M-1\right)/2$ $\Sigma$-matrices two-by-two. In the tripartite case, we label $\Sigma_{xy|\zeta}$ such that $\boldsymbol\Gamma_{xx}$ is the numerator block and $\textbf V_{y|\zeta}$ the denominator one. Given that the groups are $d_x\geq d_y\geq d_z$ from the relay, the minimum mutual information is obtained by considering groups ``Z" and ``X" and conditioning the measurement on group ``Y". The secret-key rate is then given by Eq.~(\ref{erreibgammachiegamma}) after optimizing the modulation $\mu$.

\bibliography{citations}

\end{document}